\documentclass[aps,prd,nofootinbib,psfig,graphicx,array]{revtex4}
\usepackage[colorlinks=true, pdfstartview=FitV, linkcolor=blue, citecolor=red, urlcolor=magenta]{hyperref}
\usepackage{graphicx}
\usepackage{float}
\usepackage{latexsym}
\usepackage{amsmath}
\usepackage{amsfonts}
\usepackage{amssymb}
\usepackage{verbatim}

\newcommand{\be}{\begin{equation}}
\newcommand{\ee}{\end{equation}}
\newcommand{\bea}{\begin{eqnarray}}
\newcommand{\eea}{\end{eqnarray}}

\newcommand{\ben}{\begin{eqnarray}}
\newcommand{\een}{\end{eqnarray}}

\begin{document}

\title{On global vortices in the higher derivative Lorentz-violating scenario}


\author{$^{1}$ M. Paganelly}
\email{matheuspaganelly@gmail.com}

\author{$^{2}$M. A. Anacleto}
\email{anacleto@df.ufcg.edu.br}

\author{$^{2}$F. A. Brito}
\email{fabrito@df.ufcg.edu.br}

\author{$^{2}$E. Passos}
\email{passos@df.ufcg.edu.br}

\affiliation{$^{1}$ Departamento de Física, Universidade Federal da Paraíba, Caixa Postal 5008, 58051-970, João Pessoa, Paraíba, Brazil.}

\affiliation{$^{2}$ Departamento de F\'{\i}sica, Universidade Federal de Campina Grande,
Caixa Postal 10071, 58429-900, Campina Grande, Para\'{\i}ba, Brazil.}




\begin{abstract}
We study the influence of Lorentz invariance violation (LIV) background in energy regularization of global structures in $(2, 1)$--dimensions. To this end, we consider a model in which the complex scalar and fixed three-vector couple as a high derivative order term. We show that LIV-background does not affect the energy and the equation of motion of neutral global structures. However, we observe that the charged structures are sensitive to LIV-background by presenting signatures in the electric field whose intensity is controlled by the LIV-parameter. Furthermore, the procedure developed leads to first-order solutions with finite energy and a regularized electric field.

\end{abstract}
 \maketitle


\section{Introduction}
Lorentz symmetry violation and topological defects are important issues in contemporary theoretical physics, established in the concept of symmetry breaking that play crucial roles in high-energy physics theories. The Lorentz invariance violation (LIV), anticipated in higher-dimensional theories such as string theory, introduces vacuum instabilities, potentially giving rise to nonzero expectation values for Lorentz tensorial structures, resulting in effective theories with preferred frame effects \cite{Colladay:1998fq, Kostelecky:1988zi,Filho:2022yrk}. These deviations from Lorentz invariance could manifest their effects in four-dimensional spacetime, extending the observational implications of conventional renormalizable gauge theories. Such phenomena are particularly important in cosmological evolution, where spontaneous symmetry breaking may have led to phase transitions in the early Universe, generating topological defects \cite{Vilenkin:2000jqa}. In view of studying LIV in a variety of contexts, one can address, for instance, the extension of the standard model \cite{Colladay:1998fq}, the aether field theory \cite{Carroll:2008pk}, the Horava-Lifshitz
anisotropic theory \cite{Horava:2009uw}, and the high-derivative operators \cite{Myers:2003fd}. In the present study, we shall focus on developing an investigation of the physics of topological defects in the latter LIV scenario --- see Refs.~\cite{Casana:2012ki,Belich:2006wk, Mavigno:2024bhr} for further studies in this direction.

Topological defects, such as kinks, vortices, and monopoles, can emerge in high-energy physics from the interaction between scalar and gauge fields as a result of symmetry-breaking mechanisms \cite{Manton:2004tk, Kibble:1976sj}. In particular, vortices, which arise in two-dimensional spaces, are associated with models where the scalar field couples to gauge fields through a local symmetry, forming stable, finite energy configurations \cite{Nielsen:1973cs}. This coupling provides a regularization mechanism, which allows for the formation of solutions such as the Bogomol’nyi-Prasad-Sommerfield (BPS) states, which satisfy first-order differential equations and minimize energy \cite{Bogomolny:1975de, Prasad:1975kr}. These structures can also emerge in other scenarios, such as magnetic materials \cite{Shinjo,wacho}, superfluid helium \cite{reif}, and Bose-Einstein condensates \cite{PhysRevLett.113.065302}.

On the other hand, global vortices exhibit divergent energy due to the logarithmic contribution that arises when the energy density is integrated. Nevertheless, their stability has been studied in the literature. A recent contribution \cite{Bazeia:2018wlo} has focused on global vortices, introducing a method that regularizes the energy, thus enabling new possibilities.

Building on these findings, we examined the influence of LIV-effects on the energy and equation of motion of topological defects in $(2,1)$--dimensions. The main focus is on how a model in which the LIV is incorporated via a fixed
three-vector, $u_{\mu}$, and coupled to a higher derivative operator (such as a CPT-even term obtained in \cite{deFarias:2024amf}, from dimensional projection),  impacts global vortices. 

The present work is structured as follows: In Sec.~\ref{sec2}, we explore its effect on vortex configuration and energy density, employing the Bogomol'nyi method to regularize the energy. In Sec.~\ref{charged}, we investigate charged structures under Lorentz symmetry breaking, using a non-minimal coupling between the complex scalar and gauge fields, in which the scalar field governs the electric permittivity. We also introduce a Born-Infeld contribution to analyze how an electric-field-saturating theory alters the model. Finally, in Sec.~\ref{conclu}, we conclude our analysis by discussing the results and suggesting future research directions.

\section{LIV in Global Vortices}\label{sec2}
Obeying the generic criteria to construct the LIV higher derivative operator (see Ref. \cite{Myers:2003fd}), we consider a canonical Lagrangian density of a complex scalar field $\varphi$ increased by a term with a high-order derivative in $(2,1)$ dimensions in a flat spacetime as follows:
\begin{align}
\label{liv1}
\mathcal{L} =\partial_{\mu}\Bar{\varphi}\partial^{\mu}{\varphi}+\Bar{g} (\partial_{\mu}\Bar{\varphi}) u^{\mu}u_{\nu}(u_{\alpha}\partial^{\alpha})^2(\partial^{\nu}\varphi)-V(|\varphi|,x^{2})\;,
\end{align}
where $x^{2}=x_{\mu}x^{\mu}$. In the formalism of the three-dimensional effective field action, the modifications are described by dimension five operators. In particular, $\Bar{g}$ represents a positive parameter written as $\bar{g} = \alpha M_{\rm Pl}^{-2}$, where $M_{\rm Pl}$ is the Planck mass and $u^{\mu}$ is a 3-dimensional vector, which can be either time-like, expressed in the form $u_{\mu}=(1,\vec0)$ or space-like $u_{\mu}=(0, u_{i})$ with $i=1,2$. Note that this operator is even under CPT, preserving the $\mathbb{Z}_{2}$ symmetry: $u_{\mu} \to - u_{\mu}$. 

The form of the potential in Eq.~\eqref{liv1} is an approach motivated by the Ref.~\cite{Bazeia:2003qt} that leads the original $V(|\varphi|)$ in a $V(|\varphi|,x^{2})$ breaking the translational invariance, but in return allows us to evade the Derrick-Hobart theorem and investigate time independent global structures in spatial dimensions greater than one. Recent works have presented modifications of this type to model excitations in magnetic materials with two spatial dimensions \cite{Bazeia:2016pvk, Bazeia:2016est, Bazeia:2017gue}.

The equation of motion associated with the Lagrangian density in Eq.~\eqref{liv1} is given by
\begin{align}
\partial_{\mu}\partial^{\mu}\varphi+\Bar{g}\;u^{\mu}u^{\nu}u^{\alpha}u^{\beta}\partial_{\alpha}\partial_{\beta}\partial_{\mu}\partial_{\nu}\varphi+\frac{\varphi}{2|\varphi|}V_{|\varphi|}=0\;\;.
\end{align}
The subindex $|\varphi|$ in the potential $V$ represents the derivative with respect to the absolute value of the scalar field. In order to investigate static configurations, we take $\varphi$ to be time-independent and consider the vortex ansatz
\begin{align}
    \varphi(r,\theta)=h(r)e^{in\theta}\;,
\end{align}
 with the following boundary conditions $h(0)=0$ and $ h(\infty)=a$. The angular and radial coordinates are denoted by $\theta$ and $r$, respectively, and $n$ is an integer associated with the vorticity of the solutions.

 The use of spherical coordinates allow us to write the three-vector $u_{\mu}$ in the form $u_{r}=(0,1,0)$, parallel to the radial component and $u_{\theta}=(0,0,1)$, parallel to the angular component, leading to different scenarios of LIV. Here, we focus on the case where the Lorentz symmetry is broken in the angular direction. In this way, the equation of motion takes the form
\begin{eqnarray}
    -\frac{1}{r}(rh')'+\frac{n^2 h}{r^2}\bigg(1+\frac{\Bar{g}n^2}{r^2}\bigg)+\frac{1}{2}V_{h}=0\;\;,
\end{eqnarray}
where the prime denotes the derivative with respect to the radial coordinate $r$. The above equation illustrates the standard global vortices augmented by a term $\Bar{g}n^2/r^2$ arising from the LIV. 

The energy density associated with this model is given by
\begin{align}
    \rho=h'^2+\frac{n^2h^2}{r^2}\bigg(1+\frac{\Bar{g}n^2}{r^2}\bigg)+V(h,r)\;.
\end{align}

Analyzing each term in the energy density, we note that asymptotically $h\approx a$, $h'\approx 0$, and $V(h)\approx 0$. Thus, we observe that the energy density becomes $\rho\propto\ 1/r^2$, which leads to structures with infinite energy. In order to avoid the divergence, we introduce an auxiliary function $W(|\varphi|)$ in the energy density inspired by the work \cite{Bogomolny:1975de} as follows:
\begin{align}
\label{energyd1}
    \rho=\bigg(h'-\frac{W_{|\varphi|}}{r}\bigg)^2+\frac{n^2h^2}{r^2}\bigg(1+\frac{\Bar{g}n^2}{r^2}\bigg)+V(h,r)-\frac{W^2_{|\varphi|}}{r^2}+\frac{2W'}{r}\;.
\end{align}
Thus, the energy density can be rewritten in terms of an effective potential $V_{eff}$ in the form 
\begin{align}
\label{energyd1eff}
    \rho=\bigg(h'-\frac{W_{|\varphi|}}{r}\bigg)^2+V_{eff}+\frac{2W'}{r}\;.
\end{align}
In order to eliminate the dependence of the effective potential on the winding number, we assume that the potential is given by
\begin{align}
\label{potV}
    V(h,r)=\frac{W^2_{|\varphi|}}{r^2}-\frac{n^2h^2}{r^2}\bigg(1+\frac{\Bar{g}n^2}{r^2}\bigg).
\end{align}
The procedure is inspired by the works in Refs.~\cite{Bazeia:2003qt,Bazeia:2018wlo}, which allows for a first-order formalism. The $n$-dependence of the potential reflects a phenomenological construction designed to regularize specific localized structures. Rather than restricting the theory, this approach allows the investigation of topological configurations with fixed winding numbers. Our method aims not to describe all configurations simultaneously, but to provide an effective framework to explore specific localized structures.

A closely related strategy is found in \cite{Bazeia:2023mat}, where the authors study kink-like solutions by introducing a function $f(n)$ that modifies the kinetic term of the Lagrangian density. This function depends explicitly on an integer parameter $n$, and different kink configurations emerge for each choice of $n$. These configurations are interpreted as simulating distinct types of geometric constraints, consistent with experimental observations in magnetic domain walls reported in \cite{Jubert}.

Although our work focuses on vortex structures and the 
$n$-dependence appears in the potential rather than the kinetic term, we observe a conceptual similarity: in both cases, the parameter $n$ must be tuned to produce distinct physical configurations. Therefore, the dependence on 
$n$ should not be viewed as a drawback, but rather as an essential ingredient in constructing the respective models. It provides the flexibility needed to capture nontrivial localized solutions that would otherwise be inaccessible. 

The energy density in Eq.~\eqref{energyd1eff} becomes
\begin{align}
\label{energyd12}
    \rho=\bigg(h'-\frac{W_{|\varphi|}}{r}\bigg)^2+\frac{2W'}{r}\;.
\end{align}
Integrating the energy density in Eq.~\eqref{energyd12}, we get the energy 
\begin{align}
\label{energy9}
    E=2\pi\int \bigg(h'-\frac{W_{|\varphi|}}{r}\bigg)^2 r\;dr+E_{B}\;.
\end{align}
Note that $E\geq E_B$, i.e., the energy is bounded from below by the energy $E_{B}=4\pi|W(h(\infty))-W(h(0))|$, where $2\pi$ is the solid angle in two spatial dimensions. Minimum energy solutions are achieved by eliminating the first term of the expression in Eq.~\eqref{energy9}  and satisfying the first-order equation
\begin{align}
\label{FOeq}
    h'=\frac{W_{|\varphi|}}{r}\;.
\end{align}
The procedure is able to regularize the energy since the energy density becomes $\rho=2W'/r$. On the other hand, we observe that the LIV term disappears from it and the equation of motion, leading to a scenario where the breaking of Lorentz symmetry is not noticed. Consequently, the global structures analyzed here are not influenced by Lorentz-symmetry violation.

Next, we consider a model of an electric field generated by a single point charge in a medium in which the electric permittivity is controlled by a scalar field, aiming to investigate how charged configurations can be affected by the Lorentz symmetry breaking.
\section{Charged structures in a LIV scenario}\label{charged}
The studies carried out in the previous section with global configurations motivate us to investigate how charged structures can perceive the LIV effects, since these systems involve vector fields, in addition to scalar fields, which can capture the impact of breaking Lorentz invariance. We organize this section into three parts. First, we introduce a charged structure model in which a single point charge generates an electric field inside a medium, whose electric permittivity is governed by a scalar field.
 In the second step, we present a brief overview of Born–Infeld theory in two spatial dimensions, focusing on the static case, which then serves as the foundation for the subsequent model.
Finally, in the third part, we investigate the extended model, by incorporating a Born–Infeld term.
\subsection{The first model}\label{charged1}

We start in $(2,1)$ dimensions with a Lagrangian density that involves the Maxwell terms and a complex scalar field coupled non-minimally through the generalized electric permittivity $P(|\varphi|,x^{2})$, given by 
\be\label{elemod}
\begin{aligned}
   \mathcal{L}&=-\frac{1}{4}P(|\varphi|,x^{2})F_{\mu\nu}F^{\mu\nu}+\partial_{\mu}\Bar{\varphi}\partial^{\mu}\varphi+\Bar{g} (\partial_{\mu}\Bar{\varphi}) u^{\mu}u_{\nu}(u_{\alpha}\partial^{\alpha})^2(\partial^{\nu}\varphi) -A_{\mu}j^{\mu}\;,
\end{aligned}
\ee
where $A_{\mu}$ represents the gauge field, $F^{\mu\nu}=\partial^{\mu}A^{\nu}-\partial^{\nu}A^{\mu}$ is the electromagnetic field strength tensor and $j^{\mu}$ denotes the conserved source current. The equations of motion in this case are given by
\begin{gather}
\label{eqmphi2}\partial_{\mu}\partial^{\mu}\varphi+\Bar{g}\;u^{\mu}u^{\nu}u^{\alpha}u^{\beta}\partial_{\alpha}\partial_{\beta}\partial_{\mu}\partial_{\nu}\varphi+\frac{\varphi}{8|\varphi|}P_{|\varphi|}F_{\mu\nu}F^{\mu\nu}=0,
\\
 \label{gauss}\partial_{\mu}(PF^{\mu\nu})-j^{\nu}=0\;.
\end{gather}
Here, we turn our attention to the particular case of a single point charge $e$ in the absence of currents expressed as $j^{0}=2\pi e\delta({\bf r})$ and $j^{i}=0$. The electric field in this case takes the form ${\bf E}=(F^{10}, F^{20})$. Therefore, the Gauss law in Eq.~\eqref{gauss} can be solved by
\be\label{estd}
{\bf E} = \frac{e}{P(|\varphi|,r)r}\,\hat{r},
\ee
or even more compactly ${\bf E} =E_{r}\;\hat{r}$. The trivial case of the classical vacuum, in which $P=1$, leads to a known situation, where the radial component of the electric field $E_{r} \rightarrow e/r$ diverges at the origin, $r=0$. Here, we present a procedure involving the generalized electric permittivity that can regularize this behavior. To this end, we consider the energy density of this model in the form
\begin{align}
    \rho=\frac{1}{2}P|{\bf E}|^{2}+h'^2+\frac{n^2h^2}{r^2}\bigg(1+\frac{\Bar{g}n^2}{r^2}\bigg)\;,
\end{align}
or simply
\begin{align}
    \rho=\frac{1}{2P} \frac{e^{2}}{r^{2}}+h'^2+\frac{n^2h^2}{r^2}\bigg(1+\frac{\Bar{g}n^2}{r^2}\bigg)\;.
\end{align}
One can see that $\rho \propto 1/r^2$, at the limit $r\to\infty$, which leads to solutions with infinite energy. To address this issue, we introduce an auxiliary function $W(|\varphi|)$ into the energy density as follows
\begin{align}\label{energydensity2}
    \rho=\bigg(h'-\frac{W_{|\varphi|}}{r}\bigg)^2+\frac{n^2h^2}{r^2}\bigg(1+\frac{\Bar{g}n^2}{r^2}\bigg)+\frac{1}{2P} \frac{e^{2}}{r^{2}}-\frac{W^2_{|\varphi|}}{r^2}+\frac{2W'}{r}\;.
\end{align}

In Ref.~\cite{Bazeia:2021tqt}, it was introduced a model of a global monopole with generalized electric permittivity producing a distinct behavior of the electric field for a point electric charge. More recently, in Ref.~\cite{Bazeia:2022pon}, the authors presented a medium in which the electric permittivity is governed by scalar fields that simulate a Bloch Wall, and in Refs.~\cite{Bazeia:2018eta,Bazeia:2018ykz} first-order equations were obtained for vortices and monopoles with internal structure. Motivated by these works, we make a particular choice of $P$, writing it in the form
\begin{align}
\label{permi}
    P=\frac{e^{2}}{2\bigg(W^2_{|\varphi|}-n^2h^2\big(1+\Bar{g}n^2/r^2\big)\bigg)}\;.
\end{align}
This expression follows the same line of reasoning applied to the potential in Eq.~\eqref{potV}. Minimum energy solutions can be achieved if the first-order equation $h'=W_{|\varphi|}/r$ is satisfied. In this case, the energy density is reduced to $\rho=2W'/r$, or equivalently
\begin{align}
\label{rhoH}
    \rho=2h'^2\;.
\end{align}
Note that now we have an expression for the electric permittivity in Eq.~\eqref{permi} that carries the effects of the LIV, which may influence the behavior of the electric field. To illustrate our procedure, we consider the following auxiliary function
\begin{align}
\label{W}
    W(|\varphi|)=|\varphi|^2-\frac{1}{3}|\varphi|^{3}\;.
\end{align}
Feeding the first-order equation, given in Eq.~\eqref{FOeq}, with the above auxiliary function, we get
\begin{align}
    h'=h\frac{(2-h)}{r}\;,
\end{align}
so that, after solving the differential equation, we find 
\begin{align}
\label{solu}
    h(r)=\frac{2r^2}{r^2+1}\;,
\end{align}
\begin{figure}
    \centering
    \includegraphics[width=8.2cm,trim={0cm 0cm 0 0},clip]{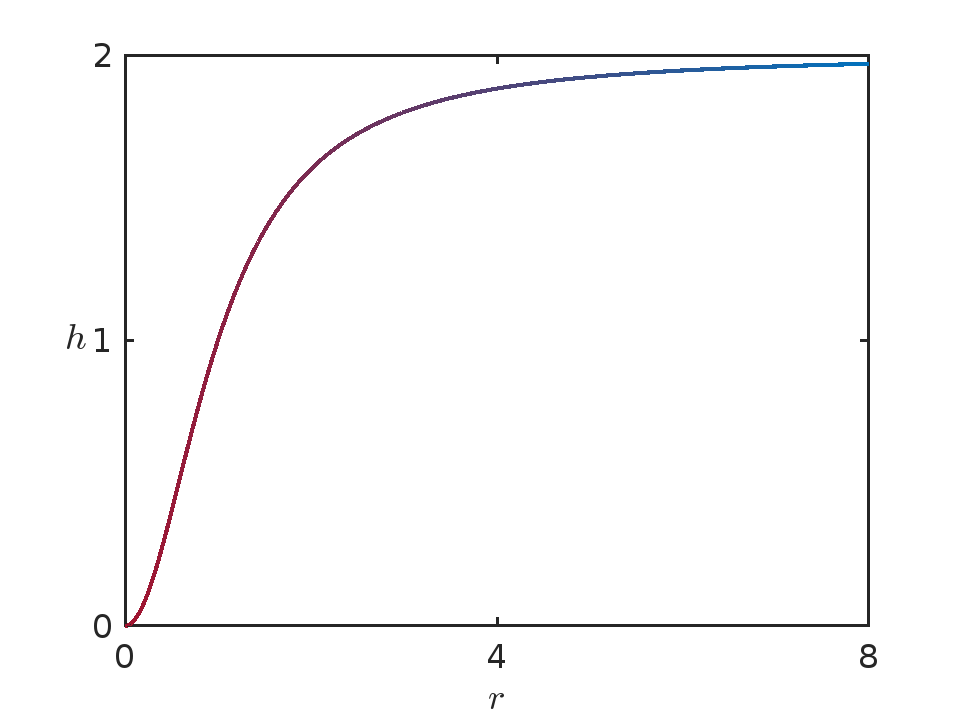}
    \includegraphics[width=8.2cm,trim={0cm 0cm 0 0},clip]{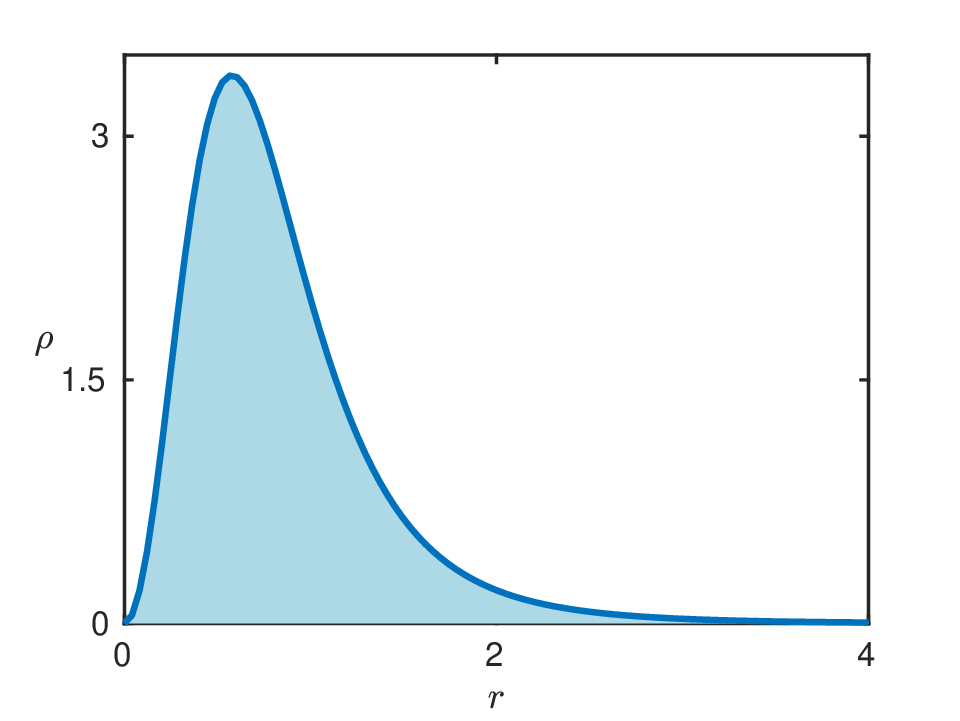}
    \caption{The solution $\phi(r)$ in Eq.~\eqref{solu} (left) and the energy density in Eq.~\eqref{rhoed} (right) associated with the model.}
    \label{hfig}
\end{figure}
and the associated energy density in Eq.~\eqref{rhoH} becomes
\begin{align}
\label{rhoed}
    \rho=\frac{32 r^2}{(r^2+1)^4}\;,
\end{align}
which remains unaffected by the constant. In Fig.~\ref{hfig}, we present both the solution and the energy density. Replacing the electric permittivity in Eq.~\eqref{permi} with the above solution in Eq.~\eqref{solu} allows us to rewrite the expression for the electric field in Eq.~\eqref{estd} as follows

\be\label{estd2}
{\bf E} = \frac{2}{er}\bigg(\frac{2r^2}{r^2+1}\bigg)^{2}\bigg[\bigg(\frac{2}{r^2+1}\bigg)^2-n^2\bigg(1+\frac{\Bar{g}n^2}{r^2}\bigg)\bigg]\,\hat{r}.
\ee

\begin{figure}
    \centering
    \includegraphics[width=8.2cm,trim={0cm 0cm 0 0},clip]{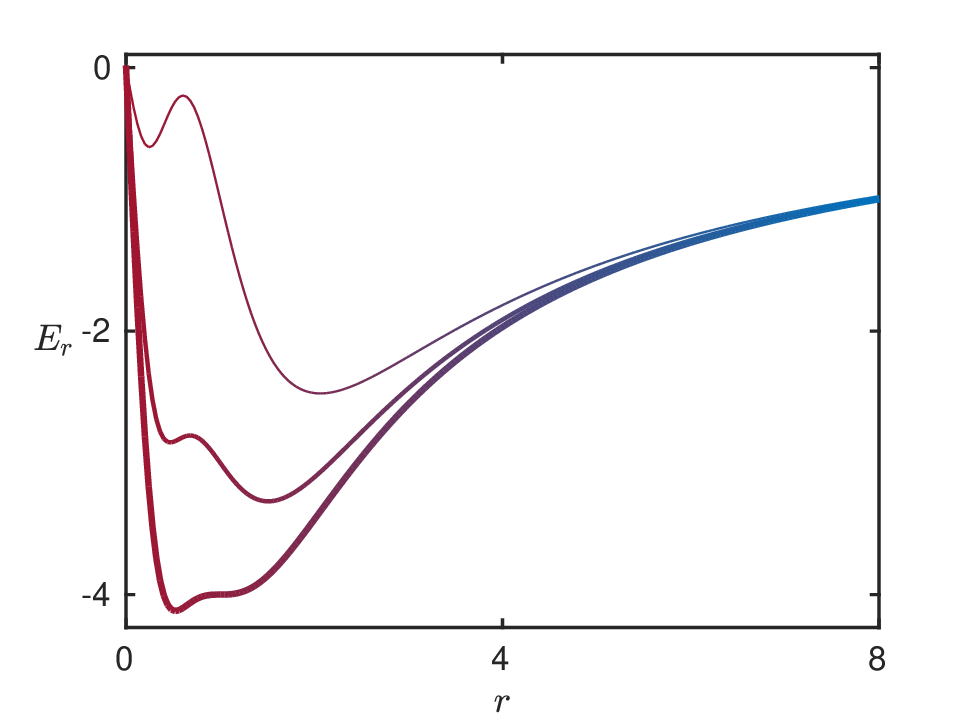}
    \caption{The radial component of the electric field in Eq.~\eqref{estd2}, with $e=1$, for $n=1$ and $\bar{g}=0.5$, $1.5$, and $2$. The thickness of
the lines increases with $\bar{g}$.}
    \label{E1fig}
\end{figure}

\begin{figure}[H]
    \centering
    \includegraphics[width=8.2cm,trim={0cm 0cm 0 0},clip]{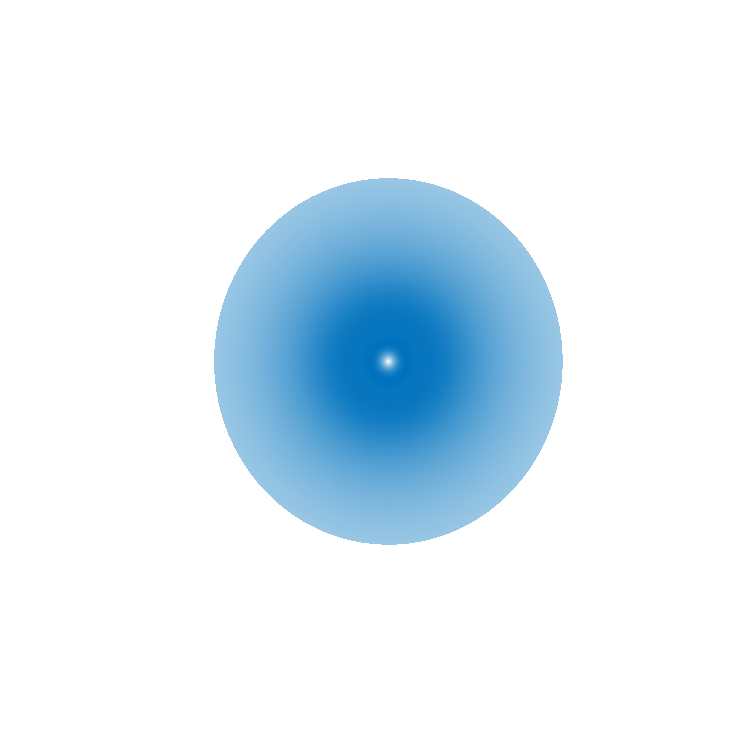} \includegraphics[width=8.2cm,trim={0cm 0cm 0 0},clip]{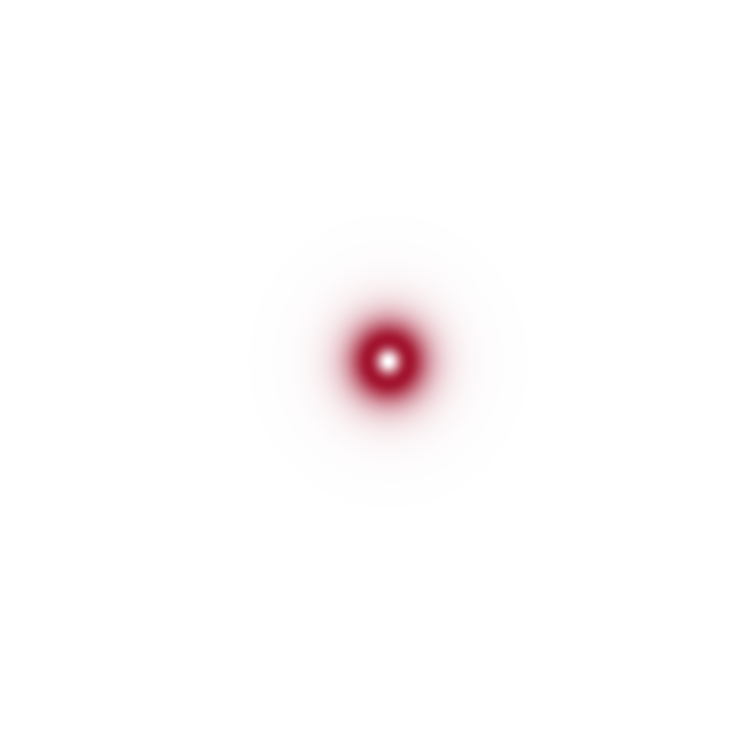}
    \caption{The radial component of the electric field in Eq.~\eqref{estd2} (left, blue) and the energy density in Eq.~\eqref{rhoed} (right, dark red) are plotted in the plane for $n=1$ and $e=1$, with $\bar{g}=2$. The intensity of the blue and dark red increases with the increase of the electric field and the energy density, respectively.}
    \label{Esurffig}
\end{figure}

The solution obtained resembles a regularized electric field typically associated with a negative charge, despite being generated by a positive particle. This effect illustrates how the electric field captures the breaking of Lorentz symmetry, which intensifies close to the origin as the parameter $\bar{g}$ increases, as shown in Fig.~\ref{E1fig}. This distinctive behavior can be interpreted as a field whose intensity is influenced by a rotating medium, enhancing its negative strength and creating a hole at the origin where the electric field vanishes. In this configuration, the scalar fields governing the electric permittivity stabilize both the electric field and the associated energy density produced by a single point charge, constraining them to a ring-shaped arrangement. To highlight these characteristics, we present the electric field and the energy density in the plane in Fig.~\ref{Esurffig}. It is worth noting that, despite these variations in field configurations, they share the same energy density, given by Eq.~\eqref{rhoed}, and with total energy $E=8\pi$. 

\subsection{A brief review of Born-Infeld theory}
In the previous section, we demonstrated that the procedure regularizes the electric field, resulting in the saturation of its intensity. A theory that exhibits similar behavior is the Born-Infeld theory \cite{Born:1934gh} (see a review in \cite{Alam:2021ovb}), which makes it an interesting study model. A model proposed in \cite{Kruglov:2016uzf}, which considers a particular case of nonlinear electrodynamics, investigated the phenomenon of vacuum birefringence, showing that the Born-Infeld theory under certain parameters nullified this effect and in addition was able to regularize the electric field of a point charge at the origin (see also \cite{Han:2016wgj}). In Ref.~\cite{Moayedi:2015lpa}, it was demonstrated that corrections to the electric field of a linearly distributed positive charge in this scenario become significant only at distances near the charge and under strong field intensities.  

We propose an extended model inspired by these works. To better contextualize our investigation, we start with reviewing the standard Born-Infeld theory for static fields. The corresponding Lagrangian density has the form
\begin{align}
\label{BI}
    \mathcal{L}&=b^2\left( 1-\sqrt{1+\frac{1}{2b^2}F_{\mu\nu}F^{\mu\nu}}\right)-A_\mu j^\mu \;,
\end{align}
where $b$ is a large constant that limits the field superiorly. The equation of motion associated with this model is given by
\begin{align}\label{eqbi}
    \partial_{\mu}\left( F^{\mu\nu}  \left(\sqrt{1+\frac{1}{2b^2}F_{\mu\nu}F^{\mu\nu}}\right)^{-1}\right)-j^{\nu}=0\;.
\end{align}
The equation above represents Gauss's law for the Born-Infeld model. The electric field that solves this equation for the case of a single point charge in the absence of currents in $(2,1)$ dimensions is given by 
\be\label{ebi}
{\bf E}_{BI} = \frac{e}{r}\frac{1}{\sqrt{1+e^2/b^2r^2}}\,\hat{r}.
\ee
One can see that the electric field is regularized and does not diverge, being limited superiorly by the constant $b$. Note that the usual electric field can be recovered if we consider that the parameter $b$ can assume an infinite value, which leads to $e^2/b^2 r^2\rightarrow 0$, and consequently ${\bf{E}}_{BI}=(e/r) \;\hat{r}$.

\subsection{The second model}

Building on this framework, we propose an extension of the model discussed in Sec.~\ref{charged1}. Here, we consider that the Born-Infeld electric field is affected by an electric permittivity controlled by scalar fields, and we investigate its behavior in the presence of Lorentz invariance violation. The corresponding Lagrangian density is given by
%
\be\label{elemod2}
\begin{aligned}
   \mathcal{L}&=b^2\left(1-\sqrt{1+\frac{P(|\varphi|,x^{2})}{2b^2}F_{\mu\nu}F^{\mu\nu}}\right)+\partial_{\mu}\Bar{\varphi}\partial^{\mu}\varphi+\Bar{g} (\partial_{\mu}\Bar{\varphi}) u^{\mu}u_{\nu}(u_{\alpha}\partial^{\alpha})^2(\partial^{\nu}\varphi) -A_{\mu}j^{\mu}\;.
\end{aligned}
\ee
The equation of motion associated with the electric field is
\begin{gather}
 \label{gauss2}\partial_{\mu}\left( P(|\varphi|,x^{2})F^{\mu\nu}  \left(\sqrt{1+\frac{P(|\varphi|,x^{2})}{2b^2}F_{\mu\nu}F^{\mu\nu}}\right)^{-1}\right)-j^{\nu}=0\;.
\end{gather}
In this case, the solution to Gauss's law is given by
\be
\label{Efinal}
{\bf \tilde{E}} = \frac{e}{P(|\varphi|,r)r}\frac{1}{\sqrt{1+e^2/P(|\varphi|)b^2 r^2}}\,\hat{r}\;,
\ee
where we use a tilde to denote the electric field associated with the aforementioned extended model. The energy density corresponding to this configuration takes the form
\begin{align}
\label{rho3}
  \rho=b^2\left(\sqrt{1+e^2/Pb^2 r^2}-1  \right)+h'^2+\frac{n^2h^2}{r^2}\bigg(1+\frac{\Bar{g}n^2}{r^2}\bigg)\;.
\end{align}
Following the procedure carried out in Eq.~\eqref{energydensity2}, we introduce an auxiliary function $W(|\varphi|)$ in the energy density given in Eq.~\eqref{rho3} and choose a particular form for the electric permittivity 
\begin{align}
  P=\frac{e^2}{b^2 r^2}\left(\Theta^2-1\right)^{-1}\;,
\end{align}
where
\begin{align}
    \Theta=\left(\frac{W_{|\varphi|}^2}{r^2}-\frac{n^2 h^2}{r^2}\left(1+\frac{\bar{g}n^2}{r^2}\right)\right)\frac{1}{b^2}+1 ,
\end{align}
which allows us to write the energy density $\rho=2W'/r$, as in Eq.~\eqref{rhoH}. 
Substituting the auxiliary function given in Eq.~\eqref{W} into the above expression leads to
\begin{align}
\label{theta}
    \Theta=\frac{1}{b^2 r^2}\bigg(\frac{2r^2}{r^2+1}\bigg)^{2}\bigg[\bigg(\frac{2}{r^2+1}\bigg)^2-n^2\bigg(1+\frac{\Bar{g}n^2}{r^2}\bigg)\bigg]+1\;.
\end{align}
The expression for $\Theta$ in Eq.~\eqref{theta} can be rewritten more compactly by using the radial component of the electric field vector from Eq.~\eqref{estd2}, obtained in Sec.~\ref{charged1} as follows
\begin{align}
    \Theta=E_r \frac{e}{2rb^2}+1\;.
\end{align}
Thus, the electric field ${\bf \tilde{E}}=\tilde{E}_{r}\;\hat{r}$ in Eq.~\eqref{Efinal} takes the form
\be\label{ef2}
{\bf \tilde{E}} = \frac{E_r}{2}\left(1+\frac{1}{E_r+2rb^2/e}\right) \;\hat{r}.
\ee

It is worth noting that the expression in parentheses in Eq.~\eqref{ef2} approaches the unit for sufficiently large values of $b$. Consequently, the electric field behaves similarly to the scenario described in Sec.~\ref{charged1} in Eq.~\eqref{estd2}, where the main difference is that the field intensity is reduced by half, i.e., it can be written as $\tilde{E}_{r} \rightarrow E_{r}$/2.

Note that the term $1/(E_{r}+2rb^{2}/e)$ in Eq. \eqref{ef2} can lead to divergences, then $b$ cannot be arbitrarily chosen. We have shown that $E_{r}$ in Eq.~\eqref{estd2} can take negative values --- see the graph in Fig.~\ref{E1fig}. Since $2rb^{2}/e$ is positive, one must select values of $b$ that cause the expression to exceed the absolute value of the minimum of $E_r$. This prevents the sum $E_{r}+2rb^{2}/e$ from disappearing, which would otherwise lead to divergences. Therefore, the following condition must be respected: $2b^{2}>-eE_{r}/r$. Since we are adopting $e=1$, we only need to take $2b^{2}>-E_{r}/r$, where this expression corresponds to the radial component in Eq.~\eqref{estd2} multiplied by the factor $-1/r$.

\begin{figure}[H]
    \centering
    \includegraphics[width=8.2cm,trim={0cm 0cm 0 0},clip]{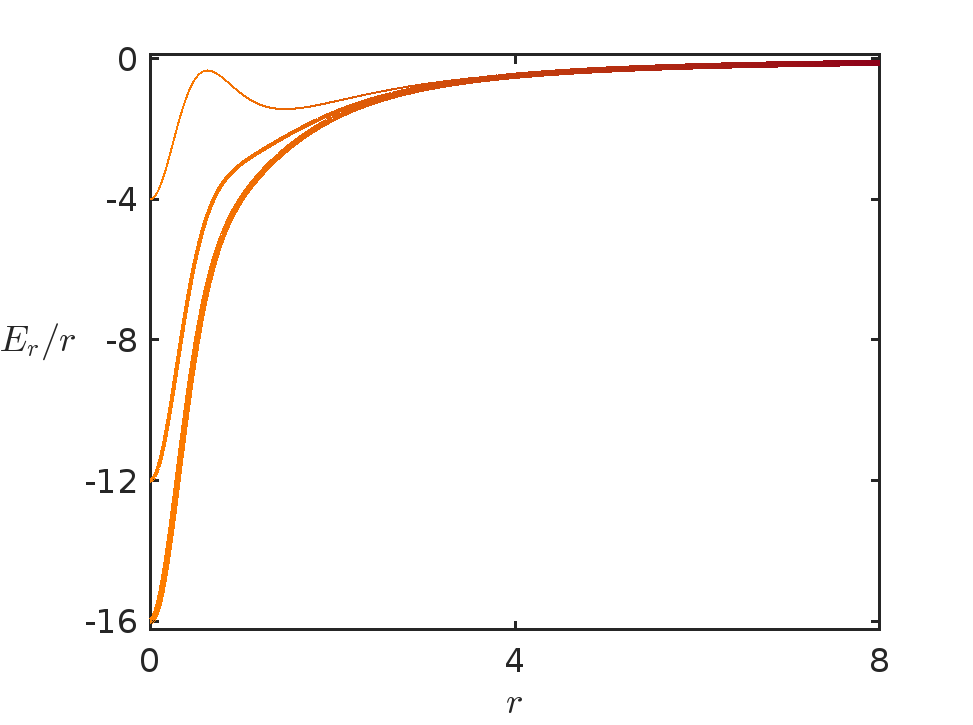}
    \caption{The radial component of the electric field in Eq.~\eqref{estd2}, multiplied by $1/r$, with $e=1$, $n=1$, and $\bar{g}=0.5$, $1.5$, and $2$, with minima  around $-4$, $-12$ and $-16$, respectively. The thickness of the lines increases with $\bar{g}$.}
    \label{comparativo}
\end{figure}

\begin{figure}[H]
    \centering
    \includegraphics[width=8.2cm,trim={0cm 0cm 0 0},clip]{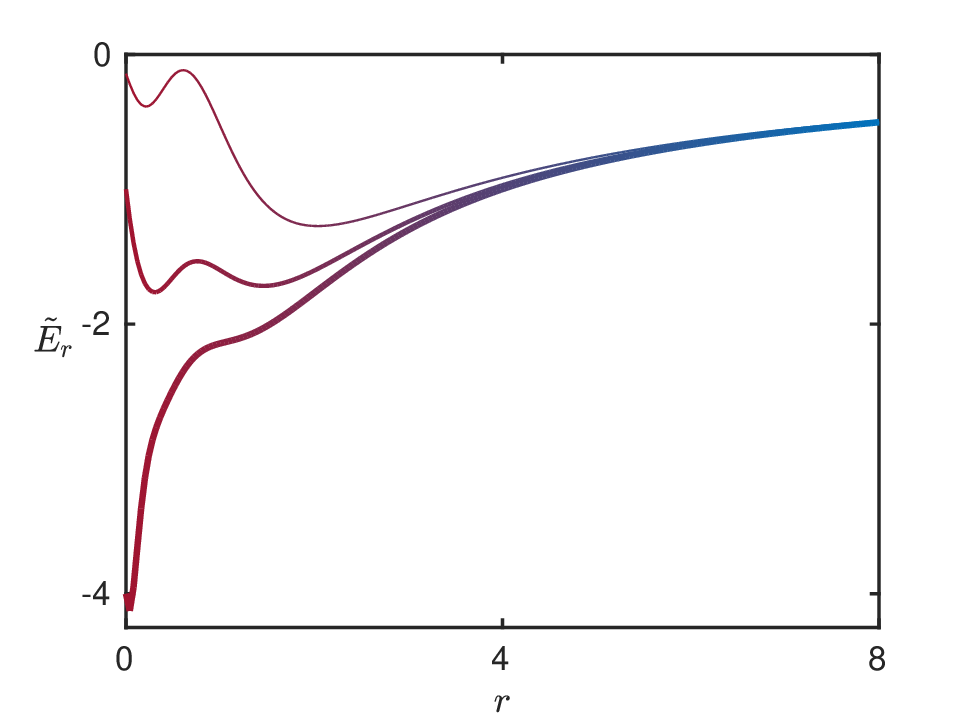}
    \caption{The radial component of electric field in Eq.~\eqref{ef2}, with $e=1$, $n=1$, and $\bar{g}=0.5$, $1.5$, and $2$ for $b=3$. The thickness of
the lines increases with $\bar{g}$.}
    \label{finalfig}
\end{figure}

\begin{figure}[H]
    \centering
    \includegraphics[width=8.2cm,trim={0cm 0cm 0 0},clip]{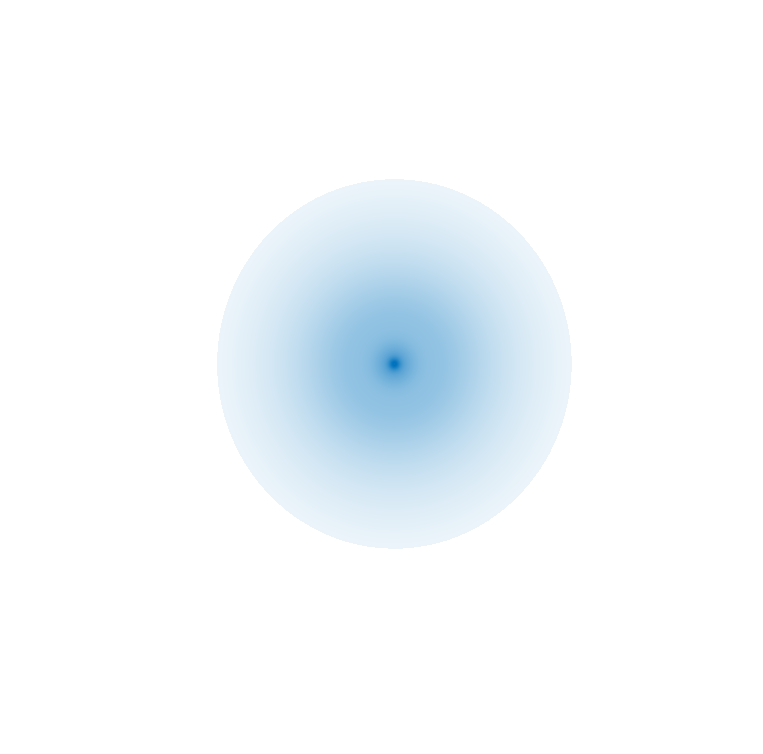}
    \caption{The radial component of the electric field in Eq.~\eqref{ef2} is plotted in the plane for $n=1$ and $e=1$ with $\bar{g}=2$ for $b=3$. The intensity of the blue color increases with the intensity of the electric field.}
    \label{E3fig}
\end{figure}

We plot $E_{r}/r$ in Fig.~\ref{comparativo}, and observe that for $g=0.5$, the expression has a minimum value around $-4$, while for $g=1.5$ and $2$, the value is close to $-12$ and $-16$, respectively. Thus, it is sufficient to take $b^{2}>8$. 
We chose the value of $b$ close to these limits and displayed the resulting configurations in Fig.~\ref{finalfig}. In doing so, we observe a distinct behavior arising from the Born-Infeld term. One notable distinction between the two models is that, in the first, in Sec.~\ref{charged1}, the electric field is zero at the origin, whereas in the second, the field reaches its highest value at the origin, although it remains finite. In the second model, the parameter $\bar{g}$, which controls the intensity of Lorentz symmetry breaking, affects the field near the origin. Although the field behaves similarly at large distances for all $\bar{g}$ values, a larger $\bar{g}$ substantially intensifies the field at the center, showing how charged structures can reflect the effects of LIV. A planar representation of the electric field is shown in Fig.~\ref{E3fig} to allow us to analyze this behavior in more detail.

\section{Conclusion}\label{conclu}
The results obtained in this study lead to advances in understanding topological defects in scenarios involving Lorentz-invariance violation (LIV). We demonstrated that for global vortices the energy regularization procedure was successful, leading to analytical solutions and eliminating the energy divergences that are characteristic of such structures. It was observed that the LIV term disappears from the equations of motion and the energy density in the case of neutral global vortices, indicating that these structures are insensitive to LIV effects.

On the other hand, when considering charged structures, the inclusion of vector fields revealed the influence of Lorentz-invariance violation. The generated electric field, although it originated from positively charged particles, exhibited a behavior analogous to that of negatively charged particles, with an intensification of the field near the origin as the LIV parameter $\bar{g}$ increases. This behavior was interpreted as an effect related to the presence of a rotating medium, showing that LIV modifies the expected behavior of the electric field near the source.

Furthermore, the inclusion of the Born-Infeld term in extended models allowed us to explore the regularization of electric field intensities in nonlinear systems. We observed that while the Born-Infeld term limits the field intensity, LIV still impacts the local configurations, demonstrating how these theories can interact to stabilize charged configurations.

Possible extensions of this work may involve investigating other kinds of topological defect under LIV conditions, such as the stable global monopoles examined in Ref.~\cite{Bazeia:2021tqt}, as well as systems incorporating electric dipoles \cite{Bazeia:2021pcw}. Furthermore, one can explore its effects in models of a bouncing Universe with non-canonical kinetic terms involving cuscutons \cite{Kim:2020iwq}. We hope that this work can serve as inspiration for future research.

{\acknowledgments} We would like to thank CNPq, CAPES and CNPq/PRONEX/FAPESQ-PB (Grant nos. 165/2018 and 015/2019),
for partial financial support. M. Paganelly acknowledges grant support 152508/2024-4. M.A.A., F.A.B. and E.P. acknowledge support from CNPq (Grant nos. 306398/2021-4,
309092/2022-1, 304290/2020-3). Furthermore, the authors extend their gratitude to P. J. Porfirio and A. Yu. Petrov for the fruitful discussions.

\end{document}